# Canonical Quantization of Neutral and Charged Static Black Hole as a Gravitational Atom

**David Senjaya[1,2] and Alejandro Saiz Rivera[1,3]**

[1]Faculty of Science, Physics Department, Mahidol University, 272 Rama VI Road, Ratchathewi District, Bangkok 10400, THAILAND

[2]anattasannya@gmail.com, [3]alejandro.sai@mahidol.ac.th

**Abstract**. The gravitational field is usually neglected in the calculation of atomic energy levels as its effect is much weaker than the electromagnetic field, but that is not the case for a particle orbiting a black hole. In this work, the canonical quantization of a massive and massless particles under gravitational field exerted by this tiny but very massive object—both neutral and charged—is carried out. By using this method, a very rare exact result of the particle's quantized energy can be discovered. The presence of a very strong attractive field and also the horizon make the energy complex valued and force the corresponding wave function to be a quasi-bound state. Moreover, by taking the small scale limit, the system becomes a gravitational atom in the sense of Hydrogenic atoms energy levels and its wave function can be rediscovered. Moreover, analogous to electronic transitions, the transition of the particle in this case emits a graviton which carries a unique fingerprint of the black hole. If it can be detected, this essential information such as black hole's mass and charge can be known.

## 1. Introduction

There is no unique 'recipe' for quantization of a classical theory. Several methods had been proposed to construct the Schrödinger equation from the classical theory. For example N T De Oliveira and R Lobo [1] found the Schrödinger equation of the classical theory of a free particle in curved space by using propagator and then expanding the corresponding action in series, taking the first two terms. De Witt [2] found the Schrödinger equation of classical theory of a free particle in curved space with certain potential energy in a quite complicated way and expressed it in kinetic term containing space metric, quantum potential containing metric and Christoffel symbol, and the potential energy term. The other approach is using Wigner function and Weyl symbol as [3].

Obtaining a quantized energy for any kind of black hole resonant system analytically is extremely rare. In this paper, that will be done via the Dirac method to construct the Hamiltonian and solve the Schrödinger-like equation—which is the quantum counterpart of the first class constraint from the classical-relativistic theory of a massive particle in curved space-time dictated by Reissner-Nordstrom (RN) and Schwarzschild metric.

## 2. Charged Black Hole and Its Canonical Quantization

The presence of a charged and extremely dense spherically symmetric static object will naturally configure the fabric of curved space-time equipped with either 2 or no horizon—which depends on the ratio between its mass and charge—called RN black hole. The metric outside that object whose mass is $M$ and total charge is $Q$—and divining $r_s = \frac{2GM}{c^2}$ also $r_Q^2 = \frac{GQ^2}{4\pi\varepsilon_0 c^4}$—reads as follows.

$$ds^2 = -\left(1 - \frac{r_s}{\tilde{r}} + \frac{r_Q^2}{\tilde{r}^2}\right)c^2 dt^2 + \frac{d\tilde{r}^2}{\left(1 - \frac{r_s}{\tilde{r}} + \frac{r_Q^2}{\tilde{r}^2}\right)} + \tilde{r}^2(d\theta^2 + \sin^2\theta \, d\phi^2) \tag{1}$$

The Lagrangian per unit mass of a particle around the RN black hole which in fact is equal with the Hamiltonian with the constrain equation reads,

$$L = \frac{1}{2}\left[-\left(1 - \frac{r_s}{r} + \frac{r_Q^2}{r^2}\right)(c\dot{t})^2 + \frac{\dot{r}^2}{\left(1 - \frac{r_s}{r} + \frac{r_Q^2}{r^2}\right)} + r^2(\dot{\theta}^2 + \sin^2\theta \, \dot{\phi}^2)\right]. \tag{2}$$

Notice that $L = \frac{1}{2}k^2c^2$ where $k^2 = 1$ for a massive particle and $k^2 = 0$ for a massless particle. It is easy to discover all of 4 conjugate momenta $p_\mu = \frac{\partial L}{\partial \dot{x}^\mu}$ and express the constraint in terms of phase space variables.

$$\Phi = \frac{1}{2}p_\mu g^{\mu\nu} p_\nu + \frac{1}{2}k^2 c^2 = 0 \tag{3}$$

Applying the canonical quantization rule to this constrain equation, we get the Schrödinger-like equation which governs the quantum dynamics of our system.

$$\left[-\frac{1}{f}\partial_{ct}^2 + \frac{1}{r^2}\partial_r(fr^2 \partial_r) + \frac{1}{r^2 \sin\theta}\partial_\theta(\sin\theta \, \partial_\theta) + \frac{1}{r^2 \sin^2\theta}\partial_\phi^2 - \frac{k^2 c^2}{\hbar^2}\right]\psi = 0 \tag{4}$$

Where $\left(1 - \frac{r_s}{r} + \frac{r_Q^2}{r^2}\right) = f = \frac{(r-r_-)(r-r_+)}{r^2}$. The temporal, azimuthal, and polar part are relatively easy to solve by the help of separation ansatz $\psi = \psi(t, r, \theta, \phi) = \tau(t)R(r)T(\theta)F(\phi)$ where $F(\phi) = e^{-im_l \phi}$, $T(\theta) = e^{i\frac{E}{\hbar c}ct}$, $T(\theta) = P_l^{m_l}(\cos\theta)$. At the end, what is left is the notoriously difficult to solve radial equation.

$$\partial_r^2 R(r) + \left[\frac{1}{(r-r_-)} + \frac{1}{(r-r_+)}\right]\partial_r R(r) + \left[\frac{E^2}{\hbar^2 c^2}\frac{r^4}{(r-r_-)^2(r-r_+)^2}\right.$$
$$\left. - \frac{l(l+1)}{(r-r_-)(r-r_+)} - \frac{E_0^2}{\hbar^2 c^2}\frac{r^2}{(r-r_-)(r-r_+)}\right]R(r) = 0 \tag{5}$$

Defining this following set of new variables,

$$\begin{aligned}\delta_r &= r_+ - r_- \\ x = r - r_+ &= \delta_r y \rightarrow dx = dr = \delta_r dy \\ r - r_- = x + r_+ - r_- &= \delta_r y + \delta_r = \delta_r(y+1) \\ \Omega &= \frac{E r_s}{\hbar c}, \quad \Omega_0 = \frac{E_0 r_s}{\hbar c},\end{aligned} \tag{6}$$

and the solution can be written in terms of the confluent Heun function [8] as $HeunC(-y) = HeunC(\alpha, \beta, \gamma, \delta, \eta, -y)$ where the parameters are $\alpha = 2\frac{\delta_r}{r_s}\sqrt{\Omega_0^2 - \Omega^2}$, $\beta = \frac{2i}{\delta_r}\frac{\Omega}{r_s}r_+^2$, $\gamma = \frac{2i}{\delta_r}\frac{\Omega}{r_s}r_-^2$, $\delta = \frac{r_+^2 - r_-^2}{r_s^2}(\Omega_0^2 - 2\Omega^2)$, $\eta = r_+^2\left(\frac{2\Omega^2 - \Omega_0^2}{r_s^2}\right) - l(l+1) - \frac{2r_+^2 r_-^2}{\delta_r^2}\frac{\Omega^2}{r_s^2}$.

$$R = e^{-\frac{1}{2}\alpha y}\left[Ay^{\frac{1}{2}\beta}(y+1)^{\frac{1}{2}\gamma} HeunC(-y)\right] \tag{7}$$

The particle's energy levels can be obtained from the condition to force the radial solution to become polynomial [4] of order $n = 1, 2, 3, \ldots$

$$\frac{\delta}{\alpha} + \frac{\beta + \gamma}{2} = -n \tag{8}$$

$$\frac{\frac{E_0 r_s}{\hbar c}\left(1 - 2\frac{E^2}{E_0^2}\right)}{2\left(1 - \frac{E^2}{E_0^2}\right)^{\frac{1}{2}}} + \frac{i}{\delta_r}\frac{E r_s}{\hbar c}(r_+^2 + r_-^2) = -n \tag{9}$$

Notice that the energy is complex valued and the imaginary part is responsible for the energy dissipation / relaxation rate and just the nature of any kind of real resonant system when it is coupled to the surrounding medium [5]. This not ever lasting bound state is called quasi bound state.

### 3. Gravitational Atom
In this section, it will be shown that by taking the limit $Er_s \to 0$ or small sized black hole we can assure the energy level to be real valued and recover the form $E_{Binding} \propto \frac{1}{n^2}$, and moreover, by taking $Q = 0$ -the outer RN horizon $r_+ \to r_s$ and inner horizon $r_- \to 0$ -which simplifies things to be Schwarzschild case.

To proceed further, define $\Omega^2 = \Omega_0^2 - \Delta\Omega^2$, $y_{Q=0} = \sigma$, $\frac{\sigma+1}{\sigma} \approx 1$, $\Omega^2\left(\frac{\sigma+1}{\sigma}\right)^2 \approx \Omega^2 + \frac{2\Omega_0^2}{\sigma}$. By this way, the radial equation is simplified a lot to be,

$$\partial_\sigma^2 R + \frac{2}{\sigma}\partial_\sigma R + \left[(\Omega^2 - \Omega_0^2) + \frac{\Omega_0^2}{\sigma} - \frac{l(l+1)}{\sigma^2}\right]R = 0. \tag{10}$$

The final substitution is $\xi = 2\sqrt{\Omega^2 - \Omega_0^2}\sigma$ and $R = \frac{1}{2\sqrt{\Omega^2 - \Omega_0^2}}\frac{S}{\xi}$ to get the famous Whittaker Hypergeometric differential equation,

$$\partial_\xi^2 S + \left[-\frac{1}{4} + \frac{\Omega_0^2}{2\sqrt{\Omega^2 - \Omega_0^2}\xi} - \frac{l(l+1)}{\xi^2}\right]S = 0$$

$$\frac{d^2 y}{dx^2} + \left(-\frac{1}{4} + \frac{k}{x} + \frac{\frac{1}{4} - p^2}{x^2}\right)y = 0 \tag{11}$$

$$y = e^{-\frac{x}{2}}\left[Ax^{\frac{1}{2}-p}{}_1F_1\left(\frac{1}{2} - k - p; 1 - 2p; x\right)\right. \\ \left. + Bx^{\frac{1}{2}+p}{}_1F_1\left(\frac{1}{2} - k + p; 1 + 2p; x\right)\right]. \tag{12}$$

The last touch for the wave function is using the identity $L_N^k(x) = \frac{\Gamma(N+k+1)}{N!\Gamma(k+1)} {}_1F_1(-N; k+1; x)$ to recover the Hydrogenic atoms radial wave function in terms of Laguerre polynomial.

$$R = C e^{-\sqrt{\Omega^2 - \Omega_0^2}\left(\frac{r}{r_s}-1\right)} \left(2\sqrt{\Omega^2 - \Omega_0^2}\left(\frac{r}{r_s}-1\right)\right)^l L_{n-l-1}^{2l+1}\left(2\sqrt{\Omega^2 - \Omega_0^2}\left(\frac{r}{r_s}-1\right)\right) \quad (13)$$

The energy quatization condition again is derived from forcing Whittaker Hypergeometric function to be polynomial of order $n = 1, 2, 3, \ldots$ by letting $\frac{1}{2} - k - p = -n$. Substituting the explicit expression of $k$ and $p$, we get

$$E_n = E_0 \sqrt{1 - \frac{G^2 M^2 m^2}{\hbar^2 c^2 n^2}} = E_0 \left(1 - \frac{G^2 M^2 m^2}{2\hbar^2 c^2 n^2} + \cdots\right). \quad (14)$$

So, for a very small mass orbiting Schwarzschild black hole, also defining $m_{pl}^2 = \frac{\hbar c}{G}$, the binding energy $E_{Binding}(n) = E_n - E_0$ will have this following form, the same as Hydrogenic atoms'.

$$E_{Binding}(n) = -\frac{mc^2}{2n^2}\left(\frac{Mm}{m_{pl}^2}\right)^2. \quad (15)$$

**Conclusion**
In summary, starting with the Lagrangian of a particle in the curved space, we have carried out the canonical quantization of a massive particle around RN black hole. We solved the Schrödinger-like equation to obtain the solution of the wave function. Then, by forcing the radial solution to be polynomial, we got the quantization rule for the complex-valued energy level (9). Taking the small black hole limit the confluent Heun series was simplified a lot to Whittaker Hypergeometric series which turned out to be just Laguerre series. Again, by forcing it to be polynomial, we got the Hydrogenic atom's energy level $E_{Binding} \propto \frac{1}{n^2}$. By this reason, this system is called gravitational atom, the atom which is governed by gravitational field. Analogously to atomic transitions emitting photons, level transitions of particles around black hole emit gravitons [6]. Anyway, experimentally, we still need next generation of instruments which is expected to achieve the necessary sensitivity to be able to detect this graviton [7].